\documentclass{iaus}
\usepackage{natbib,graphicx,wasysym}

\input{SMARTIN_SURVEYS.bibdef}
\newcommand{\mc}{\multicolumn}

\title[Extragalactic line surveys] 
{Extragalactic line surveys}

\author[Sergio Mart\'in Ruiz]   
{Sergio Mart\'in$^1$
}

\affiliation{$^1$European Southern Observatory, \\ Alonso de C\'ordova 3107, Vitacura, Casilla 19001, Santiago 19, Chile \\ email: {\tt smartin@eso.org}}

\pubyear{2011}
\volume{280}  
\pagerange{}
\jname{The Molecular Universe}
\editors{}
\begin{document}

\maketitle

\begin{abstract}
40 years have passed since the first molecular detection outside our Galaxy.
Since then, our knowledge on the distribution, kinematics and composition of the molecular material in the extragalactic ISM has built up significantly
based not only on the carbon monoxide observations but also in the more than 50 molecular species detected.
In particular, line surveys have been proven to be excellent tools to study the chemical composition in the nuclei of galaxies.
Such studies have been favored by the increasing instantaneous bandwidth of current mm and sub-mm facilities.
Here I will summarize the highlights of extragalactic molecular spectroscopy, mostly focusing in the results from molecular line surveys published in 
the last few years as well as the aims of still ongoing projects.
\end{abstract}

\firstsection 
\section{Introduction: Chemical variations in galactic nuclei}
To date, all of the spectral line surveys towards extragalactic sources have been carried out at low spatial resolution.
The large beam sizes in most of these studies just gather the averaged emission over the central few hundred parsecs in the central regions of nearby galaxies.
Almost two decades ago, the review paper by \citet{Mauersberger1993} suggested that we could naively expect that the chemical variations in the different molecular cloud complexes,
affected by different physical and chemical proccesses, would cancel out when observing the averaged emission over large regions.
However, based on a number of previous studies and the more than two dozen of extragalactic molecular detections, they claimed that
``surprisingly, the variations in the chemical composition do not cancel out'' \citep{Mauersberger1993}.
Their comparison of the molecular abundances of just seven species in the starburst galaxy prototypes, M~82 and NGC~253, where only three species were detected in both galaxies, showed that
significant differences could be identified.
Such abundance differences were attributed to the different heating mechanisms driving the chemistry in their central regions.
It became then clear the potential of chemical composition comparative studies to probe the physical processes in the heavily obscured central regions of galaxies.
Spectral line surveys, particularly at mm and submm wavelenghts, became key to understand and stablish the chemical classification and evolution of galactic nuclei.

\section{Targeted vs. Unbiased surveys}
Starburst galaxies have been the evident targets for the first molecular line surveys given the strong molecular emission.
Moreover, these galaxies are indeed the most prominent molecular emitters outside the galaxy.
The study by \citet{Wang2004} towards the central region of NGC~4945 is likely the first systematic detailed study of the chemical composition in an external galaxy.
Using the Swedish-ESO Submillimeter Telescope (SEST), they observed 80 transitions from 19 different species over a frequecy range between 82 and 354~GHz.
The derived molecular abundances and isotopic ratios measured in NGC~4945 were compared to those available in the literature from M~82 and NGC~253.
Though quite complete, this comparison was limited by the selection of the observed molecular species.
Line surveys over wide frequency ranges, on the other hand, provide a complete and unbiased description of the chemical composition in galactic nuclei.
The first unbiased spectral line survey was carried out with the IRAM~30m telescope
towards the central region of the starburst galaxy NGC~253 and covered the 2~mm atmospheric window between
129 and 175~GHz \citep{Mart'in2006}.
This scan resulted in the detection of 111 spectral features from 25 molecular species as well as three hydrogen recombination lines.
The follow up towards M~82, the brightest molecular emitter together with NGC~253, surveyed the same 2~mm window as well as the 1.2~mm band between 241 and 260~GHz (Aladro et al. Submitted).
Although expected from previous observations, the detection rate in M~82 was significantly lower, with 72 spectral features from 18 molecular species.
The combination of both line surveys resulted in the most complete comparison between the molecular composition of two extragalactic sources.
Indeed, the observed chemical differences are claimed to be directly linked to the different state of evolution of their nuclear starbursts, and therefore to the different leading heating mechanisms
affecting the ISM within their central few hundred parsecs.
While a number of species, such as CH$_3$OH, HNCO, NH$_3$, SiO, $\it NS$, $\it HOCO$, and $\it CH_2NH$,claimed to be formed in dust grains and injected into gas phase via shocks, were found to be
enhanced in NGC~253, other molecules, such as CO$^+$, HCO, HOC$^+$, $\rm c-C_3H_2$, CH$_3$CCH, and $\it NH_2CN$, mostly formed in the gas phase and likely enhanced by intense UV fields, are significantly more abundant
towards M~82.
Even though some of these differences were previously known, a number of them (in $italics$ in the list above) were pointed out from the observations covered by the line surveys.
In particular, line surveys allow the identification of particularly interesting and contrasted species such as the case of methanimine ($\it NH_2CN$) which shows rotational temperatures
similar to those measured towards the Galactic hot core Sgr~B2(M) in both M~82 and NGC~253, but shows an abundance enhancement towards the former.

\begin{table}
\begin{center}
\caption{Instantaneous bandwidth in GHz of currently available mm and submm facilities}
\label{tab.bandwidth}
\begin{tabular}{l @{\,\,\,...\,\,\,} c | l @{\,\,\,...\,\,\,}  c}
\hline 
\mc{2}{c}{Single Dish}     &  \mc{2}{c}{Interferometer} \\
\hline
MOPRA         & 8.2     &   SMA        & 8     \\
IRAM 30~m     & 8       &   CARMA      & 8     \\
Nobeyama 45~m & 32      &   PdBI       & 4     \\
APEX          & 8       &   ATCA       & 4     \\
JCMT          & 1.8     &   ALMA       & 8     \\
\hline
\end{tabular}
\end{center}
\end{table}

\section{Receiver Bandwidth Upgrades: Ongoing surveys}
\label{sec.surveys}
At that time of the first line surveys described in previous section, the available instantaneous bandwidth in both SEST and the IRAM~30m telescopes was $\sim$2~GHz.
Even with such not inconsiderable bandwidths, spectral scans projects were still significantly expensive in terms of telescope time.
To illustrate this, the surveys towards M~82 and NGC~253 each required more than one hundred hours of telescope time at the IRAM~30m telescope.
However, in the last few years, major upgrades have been performed in all the mm and submm facilities around the globe.
By the end of 2005, the MOPRA telescope commissioned the MOPS digital filterbank spectrometer which provided an $\sim8$~GHz instantaneous bandwidth.
Though unique at the time, most single-dish telescopes and interferometers catched up with similarly wide band receivers as shown in Table~\ref{tab.bandwidth}.
Just recently, the Nobeyama~45m telescope started commissioning the SAM45 spectrometer, consisting of a copy of the ALMA/ACA correlator that allows a maximum
of a 32~GHz bandwidth.

These major telescope upgrades have open the possibility of extending the exploration of wider spectral bands towards larger sample of extragalactic objects
given the significantly reduced time required for such spectral scans.
Consequently, a number of spectral line surveys have been carried out or are currently ongoing in almost every mm and submm facility.
Table~\ref{tab.Surveys} summarizes the observational details of some of the most relevant line surveys carried out at mm and submm wavelenghts.
Here we describe the main aims of theses surveys:

{\it IRAM 30m $.-$} Following the success of the 2mm line surveys of NGC~253 and M~82, a series of deep line survey follow up at 3mm are currently ongoing and near completion.
This project sample consist of 6 galaxies, including two young starbursts (M~83 and NGC~253), two evolved starbursts (M~82 and M~51), the active galactic nucleus (AGN) prototype NGC~1068 and the ULIRG Arp~220.
A total of 36 molecular species are detected in the covered band.
This study aims to set the basis of a chemical classification of the extragalactic ISM in the central few hundred parsecs of nearby bright galaxies.
The new instrumentation at the IRAM~30m telescope reduced the time required for each spectral scan by almost an order of magnitude, down to $\simeq10$~hours per source.

{\it Nobeyama 45m $.-$} Three line surveys in the 3~mm atmospheric band had just been finished at the 45~m facility making use of the recently upgraded spectrometer (Poster 2.69). 
With more than a dozen molecular species detected, these surveys will be able to establish the differences between the chemistry in starburst galaxies
(NGC~253 and IC~342) and the AGN dominated NGC~1068.
Thus the main motivation of this work is to study the chemical fingerprints of the AGN influence on the surrounding molecular gas.
The relatively small ($18''$) beam of the 45~m telescope is smaller than the starburst ring in NGC~1068, allowing
the observation of the chemical composition towards the AGN without contamination from the molecular material in starburst.

{\it APEX 12m $.-$} The spectral scans towards NGC~4945, NGC~253 and Arp~220 with the APEX telescope (Poster 2.83) are exploiting the capabilities of the FLASH receiver at APEX which processes an 8~GHz 
instantaneous bandwidth.
These surveys are exploring the higher energy transitions in the 0.9 mm atmospheric window. A total of 16 molecular species are detected at the highest frequencies ever surveyed outside the Galaxy.
Additionally, NGC~253 has been surveyed in the 1.3~mm atmospheric window from 185 to 275~GHz (Poster 2.83).
These APEX scans complete the spectral coverage towards NGC~253 over a total band of 274~GHz
between 3.5 and 0.8~mm (see Table~\ref{tab.Surveys}).

{\it SMA $.-$} The Submillimeter Array was used to carry out the first interferometric extragalactic unbiased survey (see Section~\ref{sec.arp220}). The large width of the emission lines in Arp~220 together
with the strong line confusion in this object, made it necessary to use the stable baselines achieved with the interferometer. A total of 70 spectral features from 15 molecular species
and 6 isotopologues were identified. The survey was partially confusion limited, with an average of 1.8 lines per GHz.
This spectral scan was the first carried out towards a ULIRG at a distance of 70~Mpc, an order of magnitude farther than nearby starburst such as NGC~253 and M~82.
Some of the highlights of this study are summarized in Section~\ref{sec.arp220}.

\begin{table}
\begin{center}
\caption{Extragalactic spectral scans at mm and submm wavelenghts}
\label{tab.Surveys}
\begin{tabular}{l l l l l l}
\hline 
Source                     &   Frequency Range    &  Telescope    &   Resolution        &   rms      & Reference    \\
                           &   (GHz)              &               &  (km\,s$^{-1}$)     &   (mK)     &              \\
\hline
NGC~253                    &  86-116              & IRAM~30m      & 12                  & 2          & Aladro et al. In prep.  \\
                           &  85-116              & Nobeyama~45m  & 20                  & 2-13       & Nakajima et al. In prep. \\
                           &  129-175             & IRAM~30m      & 8                   & 2-6        & \citet{Mart'in2006}  \\
			   &  185-275             & APEX~12m      & 12                  & 2          & Requena-Torres et al. In prep. \\
                           &  280-360             & APEX~12m      & 20                  & 2          & Requena-Torres et al. In prep. \\
M~82                       &  86-116              & IRAM~30m      & 12                  & 2          & Aladro et al. In prep.  \\
                           &  130-175             & IRAM~30m      & 8                   & 2-5        & Aladro et al. Submitted  \\
                           &  241-260             & IRAM~30m      & 5                   & 4-6        & Aladro et al. Submitted  \\
NGC~1068                   &  86-116              & IRAM~30m      & 12                  & 2          & Aladro et al. In prep.  \\                     
                           &  85-116              & Nobeyama~45m  & 20                  & 1-4        & Nakajima et al. In prep. \\
M~83                       &  86-116              & IRAM~30m      & 12                  & 2          & Aladro et al. In prep.  \\
M~51                       &  86-116              & IRAM~30m      & 12                  & 2          & Aladro et al. In prep.  \\
Arp~220                    &  86-116              & IRAM~30m      & 12                  & 2          & Aladro et al. In prep.  \\
                           &  202-242             & SMA           & 250                 & 4 $^1$     & \citet{Mart'in2011}  \\
                           &  280-360             & APEX~12m      & 20                  & 2          & Requena-Torres et al. In prep. \\
IC~342                     &  85-116              & Nobeyama~45m  & 20                  & 1-2        & Nakajima et al. In prep. \\
NGC~4945                   &  280-360             & APEX~12m      & 20                  & 2          & Requena-Torres et al. In prep. \\
PKS~1830-211               &  30-50               & ATCA          & 6-10                & ...        & \citet{Muller2011} \\
\hline
\end{tabular}
\end{center}
\vspace{1mm}
 {\it Notes:}\\
  $^1$ rms in mJy\,beam$^{-1}$ units 
\end{table}

\section{Going wider: high redshift wide band receivers}
In the last few years, a number of extremely wide band receivers have been developed to measure the redshift of distant heavily obscured objects.
These receiver cover wide frequency bands at a low spectral resolution. Even if this spectral resolution is similar or even wider than the tipical linewidth of 
local galaxies, it is still possible to study the brightest molecular transitions in relatively large samples of galaxies at a relatively low observing time expense.
In this section we summarize some of the latest publications making use of these wide band instruments.

{\it Redshift Search Receiver (RSR) at FCRAO 14m $.-$} The RSR, design as a facility instrument for the 50m LMT,
is a dual polarization dual beam instrument with 4 broadband receivers that covers the 37~GHz accross the 3mm atmospheric window, between 74 and 111~GHz.
The spectral resolution of $\sim100$\,km\,s$^{-1}$ provided is barely enough to resolve the typical linewidth of $\sim200$\,km\,s$^{-1}$ in the central region of local galaxies.
This spectrometer was used to survey the 3~mm atmospheric band towards a sample of 10 galaxies with different types of nuclear activity \citep[see Table 1 in][]{Snell2011}.
A total of 33 transition of 13 molecular species were detected. The low number of spectral features as compared with similar scans with the IRAM~30m or Nobeyama~45m is a consequence of 
the line smoothing due to the low spectra resolution.
Still line ratios between the brighter spectral features could be extracted from this survey.
For example the HCO$^+$/HCN ratio was observed to be enhanced in AGN dominated galaxies when compared to starburst (SB) dominated galaxies \citep{Snell2011}.
This result, in disagreement with the result from \citet{Krips2008}, where the ratio HCO$^+$/HCN is observed to be enhanced in SB galaxies, shows the difficulty of 
establishing activity templates in extragalactic sources. It is indeed difficult to disantangle the contribution from AGN and SB at the low resolution provided by single dish observations.
As such, the highest HCO$^+$/HCN ratios from the surveys from \citet{Snell2011} were derived from the LIRGs NGC~3690 and NGC~6240 where, even if known to host an active nucleus, the contribution
from the star formation cannot be excluded.

{\it Z-Spec at CSO 10.4m $.-$} Z-Spec is a millimeter wave grating spectrometer dispersing the light to an array of 160 bolometers.
Though covering an enormous bandwidth of almost 120~GHz from 190 to 307~GHz, the spectral resolution is very low.
Velocity resolution ranges from 700\,km\,s$^{-1}$ at the lower frequencies to 1200\,km\,s$^{-1}$ at the higher end.
This resolution is significantly coarser than the observed molecular linewidth even in distant galaxies.
Still, this receiver has been used to survey large frequency bands in local galaxies such as the 3 positions observed towards the starburst galaxy M~82 \citep{Naylor2010}.
With an integration of just 1 hour per position, 10 molecular species were detected within the band.
Even though the resolution was heavily limited, it is possible to simultaneous observe three transitions of CS and therefore to constrain and compare the physical properties of the molecular gas in the three
observed positions.
Similarly, \citet{Kamenetzky2011} observed the central region of NGC~1068 with Z-spec, where they identified 12 molecular species.

Though not intended as a molecular line survey, the observations with Z-Spec towards teh Cloverleaf galaxy at z$\sim$2.56 covered the rest frame wavelenghts between 272 and 444~$\mu$m \citep{Bradford2009}.
At these wavelenghts, four transition of CO (from $J=6-5$ up to $9-8$) were clearly detected and [CI] at 370~$\mu$m blended to the CO $7-6$ transition.
Additionally, tentative detection of H$_2$O in emission and CH$^+$ and LiH$^+$ in absorption were reported.
These CO simultaneous measurements, together with previous lower-$J$ detections with the IRAM PdBI, allow the detailed study of the CO spectral line energy distribution (SLED) in the cloverleaf
\citep[Figure 2 in ][]{Bradford2009}.
The studies at high-z with broad band spectrometers at ground based telescopes are similar to those carried out with Herschel space observatory toward nearby galaxies.
For example, the observations with the SPIRE spectrometer covering the frequencies from 467 to 989~GHz measured the CO emission from $J=5-4$ to $13-12$ towards the ULIRG Mrk~231 \citep{vanDerWerf2010}.
The resulting CO SLED has been modelled with a combination of dense gas exposed to a strong UV radiation and a significant X-ray contribution from the supermassive black hole to explain the 
line luminosity of the higher transitions.
Similar SLED were modelled towards M~82 based on both SPIRE and HIFI spectrometers on board Herschel \citep{Panuzzo2010,Loenen2010}.
The SPIRE spectrum towards Mrk~231 \citep{vanDerWerf2010} also resulted in the detection rotational lines of H$_2$O, OH$^+$, H$_2$O$^+$, CH$^+$, and HF, which shows the potential of molecular
studies in the submillimeter to infrared wavelenghts.

\section{Absorption systems}
\label{sec.absorp}
Among the surveys compiled in this review, those towards absorption systems deserve an special mention due to its enormous potential to carry out molecular studies at intermediate redshift.
Such is the case of the spectral scan carried out with ATCA towards the absorption system PKS~1830-211 \citep{Muller2011}.
The survey (see Invited talk in this volume by Sebastien Muller for an extensive description of the results) covered the frequency range from 30 to 50~GHz, equivalent to the $5-3$~mm rest frame between 57 and
111~GHz at the redshift of z=0.89.
As a result of the spectral scan 28 molecular species and 8 isotopologues were identified. This study turns PKS~1830-211 into one of the extragalactic sources with the largest number of molecular 
detections.
The accurate fractional abundances derived from these observations allows complete comparison with Galactic sources \citep[see Fig. 8 in][]{Muller2011} which showed the similar molecular composition
between this absorption system and that found in typical Galactic diffuse and translucent clouds.

\begin{table}
\begin{center}
\caption{Census of extragalactic molecular species and isotopologues detected}
\label{tab:census}
\begin{tabular}{ llllll }
\hline
{\bf 2 atoms}           &       {\bf 3 atoms}   &       {\bf 4 atoms}   &       {\bf 5 atoms}   &       {\bf 6 atoms}           &       {\bf 7 atoms}   \\
\hline
OH                      &       H$_2$O, {\tiny H$_2^{18}$O}
                                                &       H$_2$CO         &       c-C$_3$H$_2$    &       CH$_3$OH, {\tiny $\rm^{13}CH_3OH$} 
                                                                                                                                &       CH$_3$C$_2$H    \\
CO {\tiny \hspace{-5pt} $\Bigg\{ \hspace{-5pt} \begin{array}{l} ^{13}CO \\ C^{18}O \\ C^{17}O \\ \end{array} $}
                        &       HCN {\tiny \hspace{-5pt} $\Bigg\{ \hspace{-5pt} \begin{array}{l} H^{13}CN \\ HC^{15}N \\ DCN \\ \end{array} $}
                                                &       NH$_3$          &       HC$_3$N {\tiny \hspace{-5pt} $\Bigg\{ \hspace{-5pt} \begin{array}{l} H^{13}CCCN \\ HC^{13}CCN \\ HCC^{13}CN \\ \end{array} $}
                                                                                                &       CH$_3$CN                &       CH$_3$NH$_2$    \\
H$_2$, {\tiny $HD$}     &       HCO$^+$ {\tiny \hspace{-5pt} $\Bigg\{ \hspace{-5pt} \begin{array}{l} H^{13}CO^+ \\ HC^{18}O^+ \\ DCO^+ \\ \end{array} $}
                                                &       HNCO            &       CH$_2$NH        &                               &       CH$_3$CHO       \\
CH                      &       C$_2$H          &       H$_2$CS         &       NH$_2$CN        &                               &                       \\
CS {\tiny \hspace{-5pt} $\Bigg\{ \hspace{-5pt} \begin{array}{l} ^{13}CS \\ C^{34}S \\ C^{33}S \\ \end{array} $}
                        &       HNC {\tiny \hspace{-5pt} $\Big\{ \hspace{-5pt} \begin{array}{l} HN^{13}C \\ DNC \\ \end{array} $}
                                                &       HOCO$^+$        &       CH$_2$CO        &                               &                       \\
CH$^+$                  &       N$_2$H$^+$, {\tiny $N_2D^+$}
                                                &       C$_3$H          &       l-C$_3$H$_2$    &                               &                       \\
CN                      &       OCS             &       H$_3$O$^+$      &       H$_2$CCN        &                               &                       \\
SO, {\tiny $^{34}SO$}   &       HCO             &                       &       H$_2$CCO        &                               &                       \\
SiO, {\tiny $^{29}SiO$} &       H$_2$S          &                       &        C$_4$H         &                               &                       \\
CO$^+$                  &       SO$_2$          &                       &                       &                               &                       \\
NO                      &       HOC$^+$         &                       &                       &                               &                       \\
NS                      &       C$_2$S          &                       &                       &                               &                       \\
LiH                     &       H$_3^+$         &                       &                       &                               &                       \\
CH                      &       H$_2$O$^+$      &                       &                       &                               &                       \\ 
NH                      &       l-C$_3$H        &                       &                       &                               &                       \\  
OH$^+$                  &                       &                       &                       &                               &                       \\ 
HF                      &                       &                       &                       &                               &                       \\
SO$^+$                  &                       &                       &                       &                               &                       \\  
\hline
\end{tabular}
\end{center}
 {\it Notes:} Updated Table from \citet{Mart'in2011} with the new detections in the absorption system PKS~1830-211 by \citet{Muller2011}.\\
\end{table}

\section{The chemistry in the Arp~220 ULIRG}
\label{sec.arp220}
At a redshift of $z=0.018$, Arp~220 is the nearest ULIRG. With a star formation rate of SFR$\sim300\,M_\odot\,yr^{-1}$, this galaxy could be considered as a scaled down version of the galaxies observed
at high redshift, with SFR up to an order of magnitude higher than that in Arp~220.
The 1.3~mm line survey carried out with the SMA (Sect.~\ref{sec.surveys}) aimed to establish the differences and similarities between the molecular composition in ULIRGs and that found in local
nearby starburst galaxies. Moreover, it aimed to study whether the nuclear power source, AGN and/or SB, had an imprint in the chemical composition.
Previous to this spectral scan, a number of molecular studies had been carried out towards this source. The paper by \citet{Greve2009} presented a complete compilation of all single dish observations
of CO, HCN, HCO$^+$, HNC and HNCO, as well as CO isotopologues in Arp~220.
A detailed study of the molecular content was also carried out in absorption with the Infrared Space Observatory (ISO) between 25 to 1300~$\mu$m \citep{Gonz'alez-Alfonso2004}.

The overall chemical composition derived from the 1.3~mm band emission lines resemble that in starbursts such as NGC 253.
However, the emission from vibrationally excited transitions of HC$_3$N and CH$_3$CN, never detected in local starbursts, was detected towards Arp 220 \citep{Mart'in2011}.
Vibrational temperatures derived from these transitions are in the range of $T_{\rm vib}=300-500$\,K.
Within our Galaxy, vibrationally excited emission of HC$_3$N is observed to be tracing the hottest gas around star forming cores \citep{deVicente1997,deVicente2000}. 
In the extragalactic ISM, vibrational emission of HCN and HC$_3$N have also been recently detected towards the also heavily obscured nucleus of the LIRG NGC 4418 \citep{Costagliola2010,Sakamoto2010},
as well as $v_2=1$ $l$-type absorption lines of HCN towards Arp 220 \citep{Salter2008} and mid-infrared vibration-rotation bands of C$_2$H$_2$, HCN, and CO$_2$ in a sample of (U)LIRGs \citep{Lahuis2007}.
The high vibrational temperatures derived from these detections is suggested to be the consequence of the IR-pumping \citep{Sakamoto2010}.
Though a hot component associated to a buried AGN cannot be excluded \citep{Costagliola2010}, the detected vibrationally excited emission 
has been claimed to be most likely tracing the warmest gas in the extremely compact star forming regions within these heavily obscured environments \citep{Mart'in2011}.
This scenario is supported by the detection of the $^{18}$O isotopologue of water in the Arp~220 1.3mm line survey. The water abundance estimated from the H$_2^{18}$O/C$^{18}$O ratio of $\sim2\times10^5$ is similar
to that measured in Galactic hot cores \citep{Gensheimer1996,Cernicharo2006b}.
Such water detection should imply an enormous star formation concentration of the order of a few 10$^6$ Sgr\,B2(N)-like hot cores in a $\sim700$~pc region \citep{Mart'in2011}.

\begin{figure}
\begin{center}
 \includegraphics[width=\textwidth]{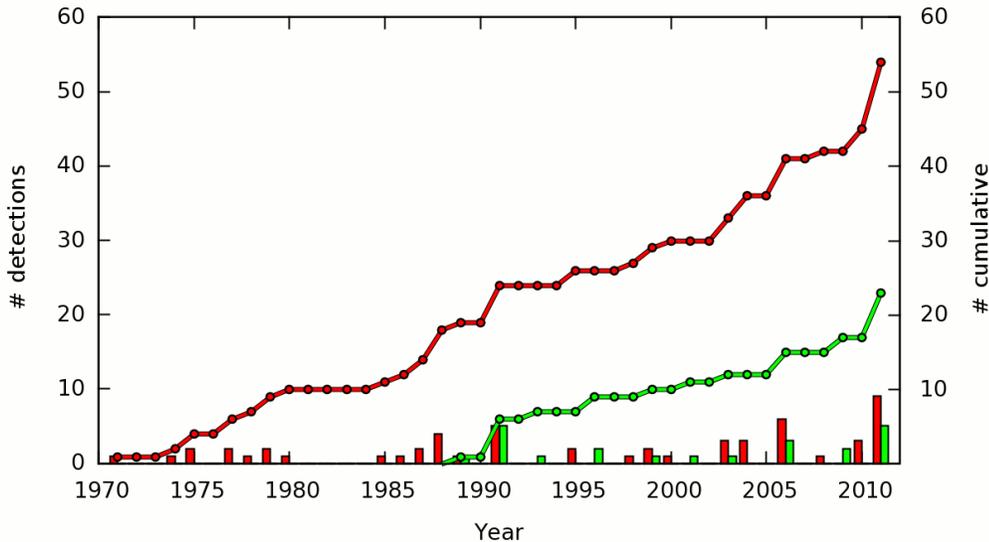} 
 \caption{Histogram shows the number of extragalactic molecular detections per year for both the main (in red) and rarer (in green) isotopologues.
The connected circles shows the cumulative number of detections.
}
   \label{fig.census}
\end{center}
\end{figure}

\section{Molecular Census Evolution}
To date, a total of 54 molecular species and 23 isotopologues have been claimed to be detected in the extragalactic interestellar medium.
Table~\ref{tab:census} shows a similar table to that presented in \citet{Mart'in2011} which became outdated by the recent detection of SO$^+$, l-C$_3$H, l-C$_3$H$_2$, H$_2$CCN, H$_2$CCO, C$_4$H, CH$_3$NH, and CH$_3$CHO
in the spectral scan towards the absorption system PKS~1830-211 (see Sect.~\ref{sec.absorp}).
It is definitely worth having a look on how the census of extragalactic molecular detections has evolved as a function of time.
Fig.~\ref{fig.census} shows the evolution with time of the cumulative number of molecular species detected in the extragalactic ISM.
Since the first molecular detection 40 years ago \citep[OH in absorption,][]{Weliachew1971} there has been a steady increase in the number of molecular detections. 
However, we can note two periods in which the rate of detections has been quicken.
First, mostly in the period between 1985 and 1990, the increase was due to the availability of sensitive radio facilities, and in particular to the large collecting area of the IRAM~30m telescope.
Only in the last few years, from 2003 up until now, spectral line scans have played the dominant role in terms of detection rate.
The advent of ALMA, with a significant increase in the collecting area of about an order of magnitud compared to the largest current facilities, will be the next step in sensitivity.
In the next decade such increase in sensitivity will allow us to match the number of extragalactic molecular detections to the number of detected species in the Galactic ISM.

\begin{figure}
\begin{center}
 \includegraphics[width=\textwidth]{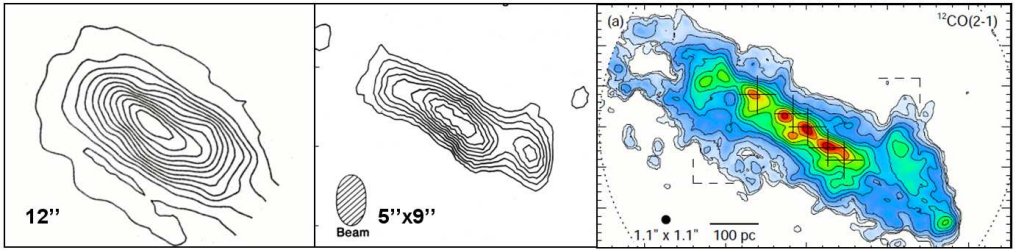} 
 \caption{
CO integrated emission maps of the starburst galaxy NGC~253 at different spatial resolutions.
(left) CO $1-0$ map at $12''$ resolution by \citet{Mauersberger1996} with the IRAM~30m telescope.
(center) CO $2-1$ map at $5''\times9''$ resolution by \citet{Canzian1988} with the OVRO interferometer.
(right) CO $1-0$ map at $1.1''$ resolution by \citet{Sakamoto2011}.
}
\label{fig.NGC253}
\end{center}
\end{figure}

\section{High Resolution Line Surveys}
Even though sensitivity is an key factor for spectral scans, the next step in molecular line surveys does not only applies to sensitivity but also to resolution.
Fig.~\ref{fig.NGC253} shows the increase in both richness and complexity of the molecular structure in the central few hundred parsecs of NGC~253 as the resolution
improves from single dish resolution to high resolution interferometric observations.
The high resolution maps by \citet{Sakamoto2011} reached a similar spatial resolution to that achieved in large scale surveys towards the Galactic center with small telescopes \citep[$\diameter=$60 cm,][]{Sawada2001}.
The imaging capabilities of instruments like ALMA, together with the large bandwidth of the receivers will allow to carry out deep 3D spectral scans.
It will be possible to study the chemical composition variations accross the center of nearby galaxies at a similar resolution to that currently achieved by large single dish telescopes towards our very Galactic center.
Thus, we expect line surveys will move a step further from studying the averaged molecular composition over the central few hundred parsecs of nearby galaxies to spatially resolve the chemical composition of individual molecular
complexes.

\bibliographystyle{aa}	
\bibliography{SMARTIN_SURVEYS.bib}

\begin{discussion}

CASELLI: You talk about ``hot core'' species. Did you try to compare abundance ratios in Arp~220 with typical ``hot core'' abundances, to test if we are indeed observing
$\sim10^6$ Sgr~B2 hot cores?.\\

MART\'IN: Such comparison was performed with the abundances measured towards the starburst galaxy NGC~253, but not yet with Arp~220.
However, Arp~220 and NGC~253 show significantly similar abundances but for the vibrational emission observed towards Arp~220.
This difference points out towards a ``hot core'' origin but a closer comparison with Galactic templates is yet to be performed.

\end{discussion}

\end{document}